\documentclass[twocolumn,showpacs,preprintnumbers,amsmath,amssymb,pre,nofootinbib]{revtex4}
\usepackage{multirow}
\usepackage{graphics}
\usepackage{array}

\newcommand{\ket}[1]{\ensuremath{\left| #1 \right\rangle}}
\newcommand{\br}[1]{\ensuremath{\left\langle #1 \right.}}
\newcommand{\bra}[1]{\ensuremath{\left. \br{#1} \right|}}
\newcommand{\bk}[2]{\br{{#1}}\ket{{#2}}}
\newcommand{\kb}[2]{\ket{{#1}}\bra{{#2}}}
\newcommand{\proj}[1]{\kb{{#1}}{{#1}}}

\newcommand{\modu}[1]{\ensuremath{\left| {#1} \right|}}
\newcommand{\magn}[1]{\ensuremath{\modu{#1}^2}}

\newcommand{\peptic}{$\Psi$-epistemic }

\begin{document}
\title{A brief note on epistemic interpretations and the Kochen-Specker theorem.}
\author{O.~J.~E.~Maroney}\email{owen.maroney@philosophy.ox.ac.uk}
\affiliation{Faculty of Philosophy, University of Oxford, 10 Merton Street, Oxford, OX1 4JJ, UK\footnote{Mailing address:  Wolfson College, Linton Road, Oxford, OX2 6UD, UK}}
\date{\today}

\begin{abstract}
One of the recent no-go theorems on \peptic interpretations of quantum proves that there are no `maximally epistemic' interpretations of quantum theory.  The proof utilises similar arrangements to Clifton's quantum contextuality proof and has parallels to Harrigan and Rudolph's quantum deficiency no-go theorem, itself based on the Kochen-Specker quantum contextuality proof.  This paper shows how the Kochen-Specker theorem can also be turned into a no `maximally epistemic' theorem, but of a more limited kind.
\end{abstract}
\pacs{03.65.Ta}

\maketitle

In \cite{Maroney2012b}, a no-go theorem is proved, regarding \peptic interpretations of quantum theory\cite{Spekkens2007,HS2007,LJBR2012}.  The theorem states that there are no `maximally epistemic' interpretations of quantum theory and that in the limit of large Hilbert spaces no more than half of the overlap between quantum states could be accounted for by epistemic uncertainty.  The proof is shown to be robust against experimental noise.  The simplest form of the proof makes use of the same experimental arrangement as Clifton's proof\cite{Clifton1993} of quantum contextuality.

A very simple proof that there are no `maximally epistemic' interpretations of quantum theory can also be obtained from the Kochen-Specker theorem\cite{KS1967}, in a manner that parallels Harrigan and Rudolph's `quantum deficiency' theorem\cite{HR2007}.  This proof is presented here.  However, unlike the theorem of \cite{Maroney2012b}, this proof does not set any bound on how epistemic a theory could become, and would not appear to be robust against finite precision loopholes.

The proof uses the ontological models framework\cite{HS2007}:
\begin{enumerate}
\item After a quantum state $\ket{\psi}$ is prepared the system is actually in a physical state $\lambda$, called the \textit{ontic} state. $\lambda$ occurs with probability $\mu_{\psi}(\lambda)$.
\[
\begin{array}{c}
\mu_{\psi}(\lambda) \ge 0 \\
\int \mu_{\psi}(\lambda) d\lambda=1
\end{array}
\]
\item A measurement procedure, $M$, has a number of possible outcomes $\{Q\}$, and has a probability, $\xi_M(Q|\lambda)$, of obtaining a particular outcome $Q$, given the ontic state $\lambda$.  The preparation procedure only influences the measurement outcomes indirectly, through the possible physical states prepared.:

     \[\begin{array}{c}
\xi_M(Q|\lambda) \ge 0 \\
\sum_Q \xi_M(Q| \lambda) =1
\end{array}\]
\item An ontological model will reproduce the results of quantum theory if, and only if:
\[
\int \mu_\psi(\lambda)\xi_M(Q|\lambda)d\lambda
=
\magn{\bk{Q}{\psi}}
 \]
\end{enumerate}

The set $\Lambda_{\phi}=\{\lambda : \mu_{\phi}(\lambda)>0\}$ represents the set of all possible ontic states which may occur when preparing the quantum state $\ket{\phi}$.  By contrast, $\Lambda^{\phi}=\{\lambda : \xi_{M}(\phi|\lambda)>0\}$ represents the set of all possible ontic states which may reveal the measurement outcome $\proj{\phi}$ when measuring $M$.  While it is clearly the case that $\Lambda_{\phi}\subseteq \Lambda^{\phi}$, the essence of Harrigan and Rudolph's theorem is that, for any ontological model for quantum theory, there must exist $\varphi$ such that $\Lambda_{\varphi}\subset \Lambda^{\varphi}$.

As $\int \mu_{\phi}(\lambda)\xi_M(\phi|\lambda)d\lambda=
\magn{\bk{\phi}{\phi}}=1$
\[
\forall \lambda \in \Lambda_{\phi} \;\;\; \xi_M(\phi|\lambda)=1
\]

If $\bk{\psi}{\phi}=0$ then
$
\int \mu_{\psi}(\lambda)\xi_M(\phi|\lambda)d\lambda
=
\magn{\bk{\phi}{\psi}}=0
$ and
\[
\forall \lambda \in \Lambda_{\psi} \;\;\; \xi_M(\phi|\lambda)=0
\]
which implies\footnote{Ignoring sets of measure zero.} $\Lambda_{\psi} \cap \Lambda_{\phi}=\emptyset$ .

If $\ket{\psi}$ and $\ket{\phi}$ are non-orthogonal, then:
\[
\int \mu_{\psi}(\lambda)\xi_M(\phi|\lambda)d\lambda
=
\magn{\bk{\phi}{\psi}} \neq 0
 \]
This allows the possibility that $\Lambda_{\psi} \cap \Lambda_{\phi}$ is not empty.

A \peptic theory is one in which there exist at least two distinct non-orthogonal quantum states, $\ket{\psi}$ and $\ket{\phi}$, for which $\Lambda_{\psi} \cap \Lambda_{\phi} \neq \emptyset$.

There is a bound on the measure of the epistemic overlap. As $\forall \lambda \in \Lambda_{\phi} \;\;\; \xi_M(\phi|\lambda)=1$
\[\begin{array}{rl}
\int_{\Lambda_{\phi}} \mu_{\psi}(\lambda) d\lambda& =\int_{\Lambda_{\phi}} \mu_{\psi}(\lambda)\xi_M(\phi|\lambda)d\lambda \\
 & \le \int \mu_{\psi}(\lambda)\xi_M(\phi|\lambda)d\lambda=\magn{\bk{\phi}{\psi}}
\end{array}\]

In a \textit{maximally} \peptic theory, for any two quantum states:
\[
\int_{\Lambda_{\phi}} \mu_{\psi}(\lambda)d\lambda
=
\magn{\bk{\phi}{\psi}}
 \]
 It is not hard to see that this will require
 \[
 \int_{\Lambda_{\phi}} \mu_{\psi}(\lambda)\xi_M(\phi|\lambda)d\lambda = \int \mu_{\psi}(\lambda)\xi_M(\phi|\lambda)d\lambda
 \]
and hence that $\Lambda_{\phi}=\Lambda^{\phi}$, showing the connection to the quantum deficiency theorem.

Now take an arbitrary state $\ket{\Psi}$ and expand it in an arbitrary basis $\{\ket{\alpha_i}\}$
\[
\ket{\Psi}=\sum_i \bk{\Psi}{\alpha_i}\ket{\alpha_i}
\]
For each overlap:
\[
\int_{\Lambda_{\alpha_i}} \mu_{\Psi}(\lambda)=\magn{\bk{\Psi}{\alpha_i}}
 \]
so (as $\bk{\alpha_i}{\alpha_j}=0$, then $\Lambda_{\alpha_i}\cap \Lambda_{\alpha_j}=\emptyset$)
\[
\int_{\cup_i\Lambda_{\alpha_i}} \mu_{\Psi}(\lambda)=\sum_i \magn{\bk{\Psi}{\alpha_i}}=1
 \]
which implies $\Lambda_{\Psi} \subseteq \cup_i\Lambda_{\alpha_i}$, up to a set of measure zero.

However all states in $\cup_i \Lambda_{\alpha_i}$ satisfy $\xi_M (\alpha_j| \lambda) \in \{0,1\}$ for any $M$.  As this must hold for arbitrary $\ket{\Psi}$ and arbitrary bases, in a maximally \peptic theory all ontic states are non-contextually value definite for all projectors.  This immediately contradicts the Kochen Specker theorem, which proves there is no model of quantum theory which is non-contextually value definite, in a Hilbert space of dimension $d \ge 3$.

Although this provides a proof that maximally \peptic theories are not possible, unlike\cite{Maroney2012b} it does not establish an empirically testable bound on the question: how close to maximally \peptic can one get?  Further, the Kochen-Specker theorem allows a finite precision loophole\cite{Meyer1999,Kent1999,CK2001,BK2004,Hermens2011}, that can be exploited to allow non-contextual theories to get arbitrarily close to quantum statistics, so it seems unlikely that this proof could be made robust against experimental error.

\textbf{Acknowledgements}
I would like to thank Chris Timpson for helpful discussions.  This research is supported by the John Templeton Foundation.

\bibliographystyle{alpha}

\end{document}